\documentclass[conference]{IEEEtran}

\usepackage{graphicx}
\usepackage[english]{babel}
\usepackage{lipsum}
\usepackage{cite}
\usepackage{xcolor}
\usepackage{dblfloatfix}
\usepackage{multirow}
\usepackage{threeparttable}
\usepackage{balance}
\usepackage{import}
\usepackage{hyperref}
\usepackage{enumitem}
\usepackage{color}
\usepackage{booktabs}
\usepackage{bm}
\usepackage{subfigure}
\usepackage{adjustbox}
\usepackage{amsmath}
\usepackage{relsize}
\usepackage[linesnumbered,vlined,plainruled]{algorithm2e}
\usepackage{array}

\definecolor{darkgreen}{rgb}{0.4, 0.6, 0.3}
\definecolor{darkgray}{rgb}{0.6, 0.6, 0.6}

\usepackage{listings}

\definecolor{codegreen}{rgb}{0,0.6,0}
\definecolor{codegray}{rgb}{0.5,0.5,0.5}
\definecolor{codepurple}{rgb}{0.58,0,0.82}
\definecolor{backcolour}{rgb}{0.95,0.95,0.95}
\definecolor{black}{rgb}{0,0,0}

\lstdefinestyle{mystyle}{
    backgroundcolor=\color{backcolour},   
    commentstyle=\color{codegreen},
    keywordstyle=\color{magenta},
    numberstyle=\tiny\color{codegray},
    stringstyle=\color{codepurple},
    basicstyle=\ttfamily\footnotesize,
    breakatwhitespace=false,         
    breaklines=true,                 
    captionpos=b,                    
    keepspaces=true,                 
    numbers=left,                    
    numbersep=5pt,                  
    showspaces=false,                
    showstringspaces=false,
    showtabs=false,                  
    tabsize=2
}

\lstdefinestyle{mystyle2}{
    backgroundcolor=\color{backcolour},   
    commentstyle=\color{codegreen},
    keywordstyle=\color{black},
    numberstyle=\tiny\color{codegray},
    stringstyle=\color{codepurple},
    basicstyle=\ttfamily\footnotesize,
    breakatwhitespace=false,         
    breaklines=true,                 
    captionpos=b,                    
    keepspaces=true,                 
    numbers=left,                    
    numbersep=5pt,                  
    showspaces=false,                
    showstringspaces=false,
    showtabs=false,                  
    tabsize=2
}

\ifCLASSINFOpdf
\else
\fi

\AtBeginDocument{\bstctlcite{BSTcontrol}}

\makeatletter
\def\@IEEEauthorblockconfadjspace{-0.7cm}
\makeatother

\begin{document}

\title{HGQ-LUT: Fast LUT-Aware Training and Efficient Architectures for DNN Inference}

\author{
    \IEEEauthorblockN{
        Chang Sun\IEEEauthorrefmark{1}\IEEEauthorrefmark{2},
        Zhiqiang Que\IEEEauthorrefmark{3}\IEEEauthorrefmark{4},
        Bakhtiar Zadeh\IEEEauthorrefmark{4},
        Qibin Liu\IEEEauthorrefmark{5},
        Kevin H. Alvarez\IEEEauthorrefmark{4},
        Wayne Luk\IEEEauthorrefmark{4},
        Maria Spiropulu\IEEEauthorrefmark{2}
    }

    \vspace{0.3cm}

    \IEEEauthorblockA{
        \IEEEauthorrefmark{2}
        California Institute of Technology, Pasadena, CA, USA,
    }
    \IEEEauthorblockA{
        \IEEEauthorrefmark{3}
        University of Bristol, UK,
    }

    \IEEEauthorblockA{
        \IEEEauthorrefmark{4}
        Imperial College London, UK,
    }
    \IEEEauthorblockA{
        \IEEEauthorrefmark{5}
        SLAC National Accelerator Laboratory, Menlo Park, CA, USA
    }

    \vspace{-1cm}
}

\newcommand{\fmax}{$\mathrm{F}_\mathrm{max}$}

\maketitle
\pagestyle{plain}
\thispagestyle{plain}

\begingroup\renewcommand\thefootnote{\IEEEauthorrefmark{1}}
\footnotetext{Corresponding author. Email: chsun@cern.ch}
\endgroup

\begin{abstract}

    Lookup-table (LUT) based neural networks can deliver ultra-low latency and excellent hardware efficiency on FPGAs by mapping arithmetic operations directly onto the logic primitives. However, state-of-the-art LUT-aware training (LAT) approaches remain difficult to use in practice: they are often orders of magnitude slower to train than conventional networks, require non-trivial manual tuning for hardware efficiency, and lack an end-to-end workflow. This work presents HGQ-LUT\footnote{Integrated in \url{https://github.com/calad0i/HGQ2}}, a new LAT approach that achieves state-of-the-art hardware efficiency while accelerating training by over 100 times on modern GPUs. HGQ-LUT introduces LUT-Dense and LUT-Conv layers that are implemented with regular, accelerator-efficient tensor operations during training, which are then compiled into logic LUTs for hardware. By combining these layers with fine-grained, element-wise heterogeneous quantization (including zero-bit pruning) and a LUT-aware resource surrogate, HGQ-LUT enables the automatic exploration of accuracy-resource trade-offs without manual bit-width tuning. We further integrate HGQ-LUT into open-source toolchains, enabling unified design, compilation, and bit-exact verification of hybrid architectures that mix LUT-based with conventional arithmetic blocks. These features make LAT-based DNNs practical for real-world deployment, such as at the CERN Large Hadron Collider's experiments.

\end{abstract}

\IEEEpeerreviewmaketitle

\section{Introduction}

Recent advances in LUT-based neural network inference have delivered remarkable hardware efficiency with ultra-low latency on Field-Programmable Gate Arrays (FPGAs), enabling high-performance deployment of Deep Neural Networks (DNNs) in resource-constrained edge and real-time computing systems. By leveraging the inherent programmability and massive parallelism of FPGA lookup-table (LUT) primitives, LUT-aware training (LAT) approaches such as LUTNet~\cite{lutnet} and the recent NeuraLUT-Assemble (NLA)~\cite{na} have shown promising results in achieving excellent hardware efficiency and ultra-low inference latency. Despite their compelling advantages, existing LAT approaches face a critical bottleneck: extremely slow training, which severely limits their practicality and scalability. For instance, current LAT methods have training times per batch that are more than two orders of magnitude slower than those of conventional neural networks. This overhead stems from complex differentiable approximations of truth tables and/or iterative reconfiguration of the connections between LUTs that require irregular memory access patterns. This limitation renders larger, more expressive LUT-aware models infeasible to train in practice, severely constraining the applicability of the LAT methods. In addition, while state-of-the-art methods like NLA~\cite{na} allow the use of mixed-precision LUT configuration to improve model expressiveness, they often remain hardware-inefficient due to the requirement of non-trivial manual configuration of bit-widths in a block-wise manner.
Moreover, current workflows rely on extensive manual intervention, custom scripts, and fragmented tools, especially for models requiring any non-LUT-based operations, such as for pre-processing, hindering rapid prototyping exploration and real-world deployment.

To address these challenges, we propose HGQ-LUT, a novel LUT-aware training framework that achieves state-of-the-art resource efficiency for FPGA-based inference while being orders of magnitude faster to optimize than prior state-of-the-art LAT methods, achieving 197$\times$ speedups over NLA~\cite{na}. HGQ-LUT introduces LUT-based dense and convolutional layers, LUT-Dense and LUT-Conv, that can be trained with standard tensor operations for efficient GPU execution and converted to logic LUTs for deployment. It combines efficient gradient surrogate techniques with scalable LUT configuration strategies, enabling fast and stable optimization without sacrificing accuracy or hardware performance. Moreover, to make the approach deployable and verifiable, we integrate HGQ-LUT into the open-source \texttt{HGQ} and \texttt{da4ml} toolchains, providing the first unified, end-to-end workflow for designing, training, and deploying hybrid neural networks that seamlessly integrate LUT-based blocks with conventional arithmetic units. This integrated framework supports co-design and verification, automated hardware mapping, and cross-layer optimization, significantly lowering the entry barrier for broader audiences.

To the best of our knowledge, this is the first work to provide an end-to-end toolchain for using LUT-based neural networks with other arithmetic-based neural network components in one unified framework, enabling hybrid architectures with both conventional and LUT-based blocks, supporting LUT-Conv operations, and multi-cycle inference of LUT-based layers with resource reuse. The main contributions of this work are:

\begin{itemize}[noitemsep,topsep=3pt]
    \item A LAT approach that achieves state-of-the-art resource efficiency for FPGA-based inference while being orders of magnitude faster to train than prior LAT methods.
    \item An end-to-end open-source workflow integrated with \texttt{HGQ} and \texttt{da4ml}, including compiler IR support for LUT operations, automated RTL generation, and bit-exact emulation for verification, with native support for hybrid LUT/arithmetic designs.
    \item A comprehensive evaluation of the proposed framework. HGQ-LUT provides significant resource and latency improvements while maintaining model accuracy compared to other LAT methods when deploying on FPGAs.
\end{itemize}

\section{Background and Related Work}
\label{sec:background}

LUT-based neural inference exploits FPGAs' $k$-input LUT primitives by replacing arithmetic with table lookups. Many studies further adopt logical LUTs (L-LUTs), i.e., learned multi-bit truth tables that are compiled onto FPGA LUT resources, enabling ultra-low latency DNN inference.

We refer to methods that replace nonlinear neuron functions by learned multi-input LUTs and optimize the LUT contents during training as LUT-aware training (LAT).  Early examples include LUTNet~\cite{lutnet} and LogicNets~\cite{logicnets}. More recent work replaces neuron computations with multi-input L-LUTs and compiles them into synthesizable logic, such as NeuraLUT~\cite{neuralut} and its extensions. NeuraLUT-Assemble (NLA)~\cite{na} improves expressiveness by assembling sub-networks into higher fan-in logical functions and adopting mixed-precision LUTs. Other efforts study decomposition and scaling, such as ReducedLUT~\cite{reducedlut}, PolyLUT / PolyLUT-Add~\cite{polylut,polylut-add}, and AmigoLUT~\cite{amigolut}. Despite good hardware efficiency, practical adoption of LAT-based approaches is still limited by training cost, manual tuning, and missing unified tooling for hybrid design and verification. This motivates HGQ-LUT, which introduces fast LUT-aware training, automatic fine-grained accuracy-resource exploration, and end-to-end integration with open-source toolchains for deployable and verifiable hybrid LUT/arithmetic designs.

Moreover, we term approaches that keep the trained network as-is but realize multiplications or matrix-vector products via memory-based lookup tables as LUT-based arithmetic mapping (LAM) methods. The LUT-LLM~\cite{he2025lut} and CD-LLM~\cite{ma2025cd} methods are examples of LAM approaches, which use LUT-based arithmetic for accelerating matmul-like operations.

\section{Design and Optimization}
\label{sec:design}

Within the LAT paradigm, we propose HGQ-LUT, a method to map DNN inference onto FPGA LUTs with the arithmetic operations predominantly mapped to logic lookup operations.

\subsection{Architecture of LUT-layers}
\label{sec:lutlayer}

While existing LAT methods, such as NLA~\cite{na}, focus on efficiency when deployed on FPGAs, they come at the expense of training time, resulting in training times that are orders of magnitude slower than those of conventional DNNs. In particular, NLA uses interconnected ensembles of logic lookup tables~\cite{na} (L-LUTs) to replace an MLP. In NLA, an L-LUT is a truth table with $F$ inputs and 1 output, with each input/output being a fixed-point number or binary vector. At training time, each L-LUT is implemented as a full-precision MLP, and the weights of the MLP are optimized via conventional backpropagation. After training, the MLP is converted to a truth table by tracing all possible inputs and outputs and then realized as a logic lookup operation in RTL. In most layers, the connections between L-LUTs are pre-defined in a tree-like fashion for reduction, while trainable mapping is used between some pre-defined blocks to recover the accuracy loss. Due to the large search space for these mappings, the authors used dynamic scatter/gather operations for the trainable mapping.

We identify two bottlenecks for training efficiency in NLA: (1) while high-fan-in L-LUTs are more expressive, they require significantly wider and deeper MLPs to approximate during training, leading to high computation costs; (2) the required dynamic scattering/gathering operations create irregular memory access patterns and suboptimal GPU utilization.

To overcome these bottlenecks, HGQ-LUT takes a different approach in the design of the LUT-based layers. For (1), we choose to use exclusively 1-input L-LUTs (not to be confused with LUT primitives). Note that the number of inputs here refers to logical inputs, not bits. A single input is likely represented by multiple bits. While 1-input LUTs are less expressive, they can be accurately approximated by a shallow MLP with only one hidden layer during training, leading to significant training speedup.

For (2), since there are only $m \times n$ possible mappings between $m$ inputs and $n$ outputs when using 1-input L-LUTs, we may realize all possible mappings simultaneously, and automatically prune away the unnecessary ones during training via quantization. At each output node of the LUT-layer, learned reduction mappings, such as the one proposed in DWN~\cite{dwn}, may be used. However, following DWN~\cite{dwn}, we found that this is not necessary to recover performance in realistic cases, whereas a simple summation is sufficient to achieve good accuracy and make training and hardware implementation significantly more efficient. Hence, we adopt summation as the reduction operation in this work. Since the mathematical structure of such a LUT-layer is similar to that of a dense layer with the summation operation as reduction, we denote such LUT-layers as \emph{LUT-Dense} layers.

The mathematical form of LUT-Dense is shown in Eq.~\eqref{eq:lutlayer}, where $x_j$ are the inputs, $a_i$ are the outputs, and $N$ is the number of inputs.
\begin{equation}
    a^{(l)}_i = \sum_{j=1}^{N} \text{L-LUT}_{i,j}(a^{(l-1)}_j)
    \label{eq:lutlayer}
\end{equation}

We further show that such LUT-layers can approximate any continuous function, similar to conventional dense layers in a multi-layer perceptron (MLP). Consider the $l$-th dense layer shown in Eq.~\eqref{eq:mlp}, where $a^{(l-1)}_j$ is the activation of the $j$-th neuron in the previous layer, $w^{(l)}_{ij}$ and $b^{(l)}_i$ are the weights and biases of the current layer, $\phi^{(l)}$ is the activation function, and $a^{(l)}_i$ is the $i$-th activation value of the current layer.
\begin{equation}
    a^{(l)}_i = \phi^{(l)}\left(\sum_{j=1}^{N} w^{(l)}_{ij} a^{(l-1)}_j + b^{(l)}_i\right)
    \label{eq:mlp}
\end{equation}
When multiple dense layers are composed, we can redefine the intermediate activations such that the nonlinearity of the previous layer is applied to the inputs of the current layer,  as shown in Eq.~\eqref{eq:mlp2}. This reparameterization does not change the function represented by the network.
\begin{equation}
    a'^{(l)}_i = \sum_{j=1}^{N} w^{(l)}_{ij} \phi^{(l-1)}\left(a'^{(l-1)}_j\right) + b^{(l)}_i
    \label{eq:mlp2}
\end{equation}

If we set $\text{L-LUT}_{i,j}(x) = w^{(l)}_{ij} \phi^{(l-1)}(x) + \frac{b^{(l)}_i}{N}$, we can fully recover the functionality of the dense layer using LUT-Dense, as Eq.~\eqref{eq:lutlayer} is now equivalent to Eq.~\eqref{eq:mlp2}. Since MLPs can approximate any continuous function by the universal approximation theorem~\cite{hornik1989multilayer}, the LUT-Dense layer can therefore also approximate any continuous function. In this way, the LUT-Dense layer can also be regarded as a direct relaxation of the dense layer, where the original point-wise activation-affine operations are replaced by more general nonlinear mappings realized by L-LUTs.

This formulation also allows for efficient GPU implementation during training. Let $x^\mathrm{in}_{b,i,j,k}$ and $x^\mathrm{in}_{b,i,j,l}$ be the $b$-th sample in the minibatch, $k$-th input and $l$-th output of one dense layer inside the MLP implementing $\text{L-LUT}_{i,j}$, we have
\begin{equation}
    x^\mathrm{out}_{b,i,j,l} = \phi\left(\sum_{k} w_{i,j,l,k}\ x^\mathrm{in}_{b,i,j,k} + b_{i,j,l}\right).
    \label{eq:mlp_lut}
\end{equation}
Assuming a contiguous memory layout, since the required indices are contiguous, the whole computation can be realized as a single, monolithic general matrix-matrix multiplication (GEMM) operation followed by an activation function, which is highly optimized on modern GPUs.

\subsection{Mixed-precision Quantization}
Naively instantiating the LUT-layer described in Section~\ref{sec:lutlayer} with dense connectivity and fixed-point quantization leads to substantial area overhead on FPGAs, as full L-LUT interconnection—analogous to dense MLP weights—is not required to recover or maintain network performance. To reduce the area overhead, we adopt HGQ's differentiable, element-wise heterogeneous quantizers~\cite{hgq} for the inputs and outputs of each L-LUT. Since 0-bit is natively supported in HGQ quantizers, pruning is automatically performed if either the input or output quantizer of an L-LUT is set to 0-bit. In particular, while it was not considered in previous works~\cite{hls4ml,na,dwn}, using the \texttt{SAT} (clamp) mode of quantizers creates considerable hardware overhead when implemented on FPGAs, since additional comparators are required for saturation. To avoid this overhead, we employ \texttt{WRAP}-mode quantizers for the inputs. Conversely, \texttt{SAT} mode is used for the outputs of each L-LUT to reduce bit width. Because each L-LUT truth table is generated offline, saturation is resolved at compile time and does not require on-chip comparator logic, resulting in no additional hardware overhead.

\section{Implementation}
\label{sec:implementation}

We build the proposed HGQ-LUT on top of the open-source \texttt{HGQ}~\cite{hgq} framework for algorithm-hardware co-design and the \texttt{da4ml}~\cite{da4ml} framework for hardware-aware optimization as well as RTL generation. In particular, we implement the LUT-layer introduced in Section~\ref{sec:lutlayer} as a new layer type in \texttt{HGQ} with optional fused batch-normalization and convolution variants. In \texttt{da4ml}, we extend the framework to support logic lookup operations, along with the corresponding parsers and RTL generators for the LUT-layers.

\begin{algorithm}[t]
    \KwData{$X_\mathrm{in}$: the input tensor, shape $(..., C_\mathrm{in})$\\
        $L_h$: number of hidden layers used in the MLPs realizing the L-LUTs\\
        $q_\mathrm{in}, q_\mathrm{out}$: the input/output quantizers of the L-LUTs\\
        $W[L_h+1]$: list of weights of the MLPs\\
        $B[L_h+1]$: list of biases of the MLPs \\
        $\sigma$: the activation function used in the MLPs \\
        use\_batchnorm: whether to fuse batch-normalization into the LUT-Dense
    }
    \KwResult{$X_\mathrm{out}$: the output tensor, shape $(..., C_\mathrm{out})$}
    $X'_\mathrm{in} \gets \mathrm{broadcast}(X_\mathrm{in}, \text{shape}=(..., C_\mathrm{in}, C_\mathrm{out}, 1))$\;
    $X \gets q_\mathrm{in}(X'_\mathrm{in})$\;

    \For{$i$ in $L_h$}{
        $X \gets$ {einsum(...iod, iode $\rightarrow$ ...ioe, $X$, $W[i]$)}\;
        $X \gets \sigma(X + B[i])$\;
    }
    $X \gets$ einsum(...iod, iod $\rightarrow$ ...io, $X$, $W[-1]$) $+B[-1]$\;

    \If{$\mathrm{use\_batchnorm}$}{
        $X \gets$ batchnorm($X$, axis=$-1$)\;
    }

    $X^q \gets q_\mathrm{out}(X)$\;
    $X_\mathrm{out} \gets$ sum($X^q$, axis=$-2$)\;
    \Return{$X_\mathrm{out}$}\;
    \caption{GPU-friendly LUT-layer Implementation}\label{alg:lutlayer}
\end{algorithm}

\subsection{Algorithm-Hardware Co-design with HGQ}

We implement the LUT-Dense as a new layer type in \texttt{HGQ} using einsum operations for efficient GPU implementation with multidimensional tensor support. Given an input tensor $X_\mathrm{in}$ of shape $(..., C_\mathrm{in})$ and an output tensor $X_\mathrm{out}$ of shape $(..., C_\mathrm{out})$, where the ellipsis denotes arbitrary leading dimensions, the LUT-layer can be implemented as a series of einsum operations as shown in Algorithm~\ref{alg:lutlayer}. In practice, we find that MLPs implementing the L-LUTs typically require only a single hidden layer with a \texttt{tanh} activation function. As this implementation avoids random memory access patterns and leverages XLA compilation~\cite{xla_compilation}, training speed is significantly improved compared to prior LUT-aware training methods. In the algorithm, $q_\mathrm{in}$ and $q_\mathrm{out}$ are the input and output HGQ quantizers with element-wise trainable bit-widths. $W$ and $B$ are lists of weights and biases of the MLPs realizing the L-LUTs, and $\sigma$ is the activation function used in the MLPs. When \texttt{use\_batchnorm} is set to True, batch-normalization is applied before the output quantization. The final output is obtained by summing over the quantized outputs of the MLPs, or equivalently, summing over the outputs of the L-LUTs in the hardware implementation. With the support of multidimensional tensors of the LUT-Dense layer, we further implement the LUT-based convolutional layer, \emph{LUT-Conv}, with an im2col~\cite{im2col} operation preceding the LUT-Dense layer.

Since HGQ relies on a differentiable resource surrogate, Effective Bit Operations (EBOPs)~\cite{hgq}, for automatic quantization, we further extend the EBOPs formulation to support LUT-Dense. When the HGQ models are implemented with \texttt{da4ml}, we empirically observe that $\exp(0.985\cdot\log(\mathrm{EBOPs}))\approx$ \#$\mathrm{LUTs}$. At the per-layer level, we approximate the resource usage as the number of LUTs.

Consider an L-LUT with an $m$-bit input and an $n$-bit output implemented using LUT-$X$ primitives, where $X$ denotes the size of the FPGA LUT primitive. When $m \ge X$ and sufficient built-in multiplexers are available (e.g., F7/F8 in Xilinx 7-series devices or F9 in newer architectures), the implementation requires $2^{m-X} \times n$ LUT-$X$ units. When each of the LUT-$X$ primitive supports splitting into $2^{X-Y}$ smaller LUT-$Y$'s, the condition can be further relaxed to $m \ge Y$. When $m < Y$, since the algorithm used in the backend synthesis tools to pack small LUTs is unknown, we heuristically approximate the number of LUT-$X$s required as $m/Y \times 2^{Y-X} \times n$. Hence, the total number of LUTs required for one L-LUT can be approximated as shown in Eq.~\eqref{eq:luts_lut}.

\vspace{-0.2cm}
\begin{equation}
    \mathrm{EBOPs}_\mathrm{L\textrm{-}LUT} =
    \begin{cases}
        2^{m-X} \times n,                    & m \ge Y \\
        \frac{m}{Y} \times 2^{Y-X} \times n, & m < Y
    \end{cases}
    \label{eq:luts_lut}
\end{equation}

\subsection{Custom Instruction and RTL Generation}

Since \texttt{da4ml}'s internal representation, distributed arithmetic instruction set (DAIS) does not natively support logic lookup operations, we extend DAIS to include a new instruction type, \texttt{L-LUT}, to represent the L-LUT operations with an auxiliary truth-table attribute attached to the DAIS program.
During the RTL generation phase, the tables are parsed and realized as logic lookup operations in Verilog or VHDL. In the \texttt{da4ml} frontend parser, the LUT-layers defined in \texttt{HGQ} are parsed and converted into a series of \texttt{L-LUT} instructions with the corresponding truth tables extracted from the trained model, along with the necessary quantization and addition/subtraction instructions.

To further improve the ease of use of the framework, we also extend the DAIS interpreter to support the LUT operations, allowing the user to simulate the trained models in a strict bit-exact fashion (up to 64 bits internally) on CPUs before the behavioral simulation of the generated RTL projects.

Truth table generation is performed by enumerating all possible input combinations of each L-LUT, passing them through the MLPs realizing the L-LUTs, and quantizing the outputs with the output quantizer. To speed up the process, all L-LUTs in the same LUT-layer with the same input width are processed in parallel within the same Einsum operations. The conversion time of a LUT-layer with 32 inputs and 32 outputs is typically around 100 ms on a modern CPU.

Since the LUT-layers and the corresponding \texttt{L-LUT} instructions are fully embedded in the HGQ and \texttt{da4ml} toolchains, all other existing operations, such as the matmul-based dense layers or convolutional layers, can be seamlessly mixed with the LUT-layers to support hybrid architectures. By defining the model with a mixture of conventional matmul-based layers and LUT-layers, larger models can be supported while still benefiting from the low-latency inference of the LUT-layers, and the RTL conversion and hardware deployment flow remains exactly the same.

\begin{figure}[t]
    \centering
    \includegraphics[width=0.48\textwidth]{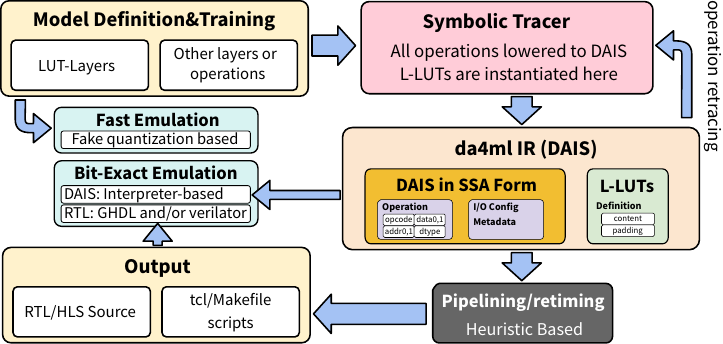}
    \caption{The overall workflow of the HGQ-LUT framework, with the LUT-layer natively supported in both \texttt{HGQ} and \texttt{da4ml}.}
    \label{fig:workflow}
    \vspace{-0.2cm}
\end{figure}

Both LUT-Dense and LUT-Conv layers are natively supported in the \texttt{HGQ} and \texttt{da4ml} toolchains, enabling easy integration of LUT-based layers with other arithmetic-based layers in one unified workflow. The overall workflow is shown in Fig.~\ref{fig:workflow}: after the user first trains and validates the model in \texttt{HGQ}, the trained model is then passed to \texttt{da4ml} to be lowered into DAIS, and then emitted as Verilog or VHDL for hardware synthesis. Bit-exact simulation is supported at both the DAIS (no compilation required) and RTL levels (requires compilation with GHDL~\cite{ghdl} and/or Verilator~\cite{verilator}) for functional verification.

\section{Evaluation}

\subsection{Experimental Setup}

\begin{table}
    \centering
    \caption{
        Training time per batch (millisecond) of HGQ-LUT and others.
    }
    \vspace{-0.2cm}
    \label{tab:training_speed}
    \begin{adjustbox}{width=0.48\textwidth}
        \begin{threeparttable}
            \begin{tabular}{llccccc}
                \toprule
                Task              & Batch size & HGQ-LUT  & HGQ   & Keras & NLA  & KANELE \\
                \midrule
                JSC HLF           & 16600      & 0.833    & 0.645 & 0.414 & 164. & 112.   \\
                JSC PLF (P32F16)  & 2790       & 1.79     & 1.85  & 1.25  & -    & -      \\
                JSC PLF (P64F16)  & 2790       & 3.14     & 4.10  & 2.76  & -    & -      \\
                TGC Muon Tracking & 51200      & 7.88$^*$ & 2.30  & 1.68  & -    & -      \\
                \bottomrule
            \end{tabular}
            \footnotesize
            $^*$ Results reproduced using hybrid architecture with both LUT-Dense and matmul-based layers.
        \end{threeparttable}
    \end{adjustbox}
    \vspace{-0.5cm}
\end{table}

\begin{table*}
    \centering

    \caption{Performance and resource consumption of the HLF JSC models.}
    \vspace{-0.2cm}
    \label{tab:hlf_jsc}

    \begin{adjustbox}{width=0.85\textwidth,center=\textwidth}

        \begin{threeparttable}
            \begin{tabular}{l|ccccccc}

                \multicolumn{8}{c}{\textbf{HLF JSC (OpenML)}}                                                                                                           \\
                \midrule
                Implementation                                      & Accuracy $\uparrow$ & Latency [cycles] & LUT           & DSP & FF     & \fmax [MHz] & II [cycles] \\
                \midrule
                \textbf{HGQ-LUT}                                    & 76.9\%              & 6 (9.2 ns)       & 5,667         & 0   & 2,434  & 649.8       & 1           \\
                \textbf{HGQ-LUT}                                    & 76.5\%              & 5 (6.5 ns)       & 2,295         & 0   & 1,145  & 771.0       & 1           \\
                \textbf{HGQ-LUT}                                    & 76.0\%              & 5 (6.4 ns)       & 1,293         & 0   & 822    & 777.0       & 1           \\
                \textbf{HGQ-LUT}                                    & 75.3\%              & 4 (4.7 ns)       & 616           & 0   & 407    & 851.8       & 1           \\
                \textbf{HGQ-LUT}                                    & 74.2\%              & 3 (3.0 ns)       & 226           & 0   & 179    & 984.3       & 1           \\
                \texttt{HGQ} [FPGA'26]~\cite{hgq}                   & 76.9\%              & 20 (36.3 ns)     & 10,182        & 0   & 10,480 & 551.6       & 1           \\
                \texttt{HGQ} [FPGA'26]~\cite{hgq}                   & 75.6\%              & 12 (18.6 ns)     & 2,298         & 0   & 2,217  & 645.2       & 1           \\
                \texttt{HGQ} [FPGA'26]~\cite{hgq}                   & 73.2\%              & 6 (9.1 ns)       & 366           & 0   & 363    & 662.7       & 1           \\
                \texttt{QKeras} [ICFPT'23]~\cite{dsp-prune}$^b$     & 76.3\%              & 15 (105.0 ns)    & 5,504         & 175 & 3,036  & $>142.9$    & 2           \\
                \texttt{DWN} [ICLR'24]~\cite{dwn}$^c$               & 76.3\%              & 10 (14.4 ns)     & 6,302         & 0   & 4,128  & 695.        & 1           \\
                \texttt{QKeras} [CoRR'21]~\cite{hls4ml}$^b$         & 76.2\%              & 9 (45 ns)        & 63,251        & 38  & 4,394  & $\sim 200$  & 1           \\
                \texttt{MetaML-Pro} [TRETS'26]~\cite{metamlpro}$^b$ & 76.1\%              & 10 (50 ns)       & 13,042        & 70  & N/A    & $\sim200$   & 1           \\
                \texttt{KANELE} [FPGA'26]~\cite{kanele}$^a$         & 76.0\%              & 7 (7.1 ns)       & 1,232 (1,488) & 0   & 900    & 987.        & 1           \\
                \texttt{TreeLUT} [FPGA'25]~\cite{treelut}           & 75.6\%              & 2 (2.7 ns)       & 2,234         & 0   & 347    & 735.        & 1           \\
                \texttt{NLA} [FCCM'25]~\cite{na}$^a$                & 75.5\%              & 2 (3.6 ns)       & 2,036         & 0   & 420    & 558.0       & 1           \\
                \midrule
                \multicolumn{8}{c}{\textbf{HLF JSC (CERNBox)}}                                                                                                          \\
                \midrule
                Implementation                                      & Accuracy $\uparrow$ & Latency [cycles] & LUT           & DSP & FF     & \fmax [MHz] & II [cycles] \\
                \midrule
                \textbf{HGQ-LUT}                                    & 75.4\%              & 6 (10.1 ns)      & 6,042         & 0   & 2,438  & 592.8       & 1           \\
                \textbf{HGQ-LUT}                                    & 75.1\%              & 6 (9.0 ns)       & 3,391         & 0   & 1,675  & 663.1       & 1           \\
                \textbf{HGQ-LUT}                                    & 74.5\%              & 5 (6.0 ns)       & 1,435         & 0   & 903    & 833.3       & 1           \\
                \textbf{HGQ-LUT}                                    & 73.7\%              & 4 (4.4 ns)       & 666           & 0   & 478    & 914.9       & 1           \\
                \textbf{HGQ-LUT}                                    & 73.0\%              & 4 (4.2 ns)       & 460           & 0   & 376    & 951.5       & 1           \\
                \texttt{HGQ} [FPGA'26]~\cite{hgq}                   & 75.3\%              & 18 (31.1 ns)     & 10,921        & 0   & 11,183 & 578.4       & 1           \\
                \texttt{HGQ} [FPGA'26]~\cite{hgq}                   & 74.5\%              & 13 (20.4 ns)     & 3,152         & 0   & 2,941  & 636.9       & 1           \\
                \texttt{HGQ} [FPGA'26]~\cite{hgq}                   & 72.4\%              & 9 (9.9 ns)       & 623           & 0   & 642    & 905.8       & 1           \\
                \texttt{KANELE} [FPGA'26]~\cite{kanele}$^a$         & 75.1\%              & 7 (8.1 ns)       & 5,034 (5,318) & 0   & 1,917  & 870.        & 1           \\
                \texttt{PolyLUT} [TC'25]~\cite{polylut}             & 75.1\%              & 5 (24.6 ns)      & 246,071       & 0   & 12,384 & 203.        & 1           \\
                \texttt{NLA} [FCCM'25]~\cite{na}$^a$                & 74.9\%              & 7 (10.3 ns)      & 8,819         & 0   & 2,770  & 679.8       & 1           \\
                \texttt{PolyLUT-Add} [FPL'24]~\cite{polylut-add}    & 75.\%               & 5 (15.9 ns)      & 36,484        & 0   & 1,209  & 315.        & 1           \\
                \texttt{NeuraLUT} [FPL'24]~\cite{neuralut}          & 75.\%               & 5 (13.6 ns)      & 92,357        & 0   & 4,885  & 368.        & 1           \\
                \texttt{ReducedLUT} [FPGA'25]~\cite{reducedlut}     & 74.9\%              & N/A              & 58,409        & 0   & N/A    & 302.8       & N/A         \\
                \texttt{QKeras} [NMI'21]~\cite{qkeras}$^b$          & 74.8\%              & 11 (55 ns)       & 39,782        & 124 & 8,128  & $\sim 200$  & 1           \\
                \texttt{QKeras} [NMI'21]~\cite{qkeras}$^b$          & 72.3\%              & 11 (55 ns)       & 9,149         & 66  & 1,781  & $\sim 200$  & 1           \\
                \texttt{AmigoLUT} [FPGA'25]~\cite{amigolut}         & 74.4\%              & 5 (9.6 ns)       & 42,742        & 0   & 4,717  & 520.        & 1           \\
                \texttt{AmigoLUT} [FPGA'25]~\cite{amigolut}         & 72.9\%              & 5 (5.0 ns)       & 1,243         & 0   & 1,240  & 1,008.      & 1           \\
                \texttt{LogicNets} [FPL'20]~\cite{logicnets}        & 71.8\%              & 5 (11.7 ns)      & 37,931        & 0   & 810    & 427.        & 1           \\
                \midrule
            \end{tabular}
            \vspace{-0.15cm}
            \footnotesize
            $^a$ Results reproduced/corrected due to the issues mentioned in Section~\ref{sec:hlf_jsc}. \\
            $^b$ \fmax is not provided, best estimated based on the target frequency/mentioning in the original works. \\
            $^c$ The results are reproduced in ref.~\cite{na} addressing the pre-processing issues. \\
            \normalsize
        \end{threeparttable}
    \end{adjustbox}

    \vspace{-0.8cm}
\end{table*}
\begin{table*}
    \centering
    \caption{
        Performance and resource consumption of the PLF JSC models and muon tracking models on FPGA.
    }
    \vspace{-0.2cm}
    \label{tab:plf_jsc_tgc}

    \begin{adjustbox}{width=0.85\textwidth,center=\textwidth}
        \begin{threeparttable}
            \begin{tabular}{lc|ccccccc}
                \multicolumn{8}{c}{\textbf{PLF JSC (3 features)}}                                                                                                  \\
                \midrule
                Implementation                     & Particles & Accuracy $\uparrow$ & Latency [cycles] & LUT       & DSP    & FF      & \fmax [MHz] & II [cycles] \\
                \midrule
                \textbf{HGQ-LUT (GNN)}             & 32        & 78.3\%              & 11 (29.9 ns)     & 41,594    & 0      & 16,402  & 367.9       & 1           \\
                \texttt{HGQ (GNN)}                 & 32        & 79.2\%              & 24 (108.8 ns)    & 159,238   & 0      & 91,027  & 220.5       & 1           \\
                \texttt{HGQ (GNN)}                 & 32        & 78.5\%              & 20 (72.3 ns)     & 80,618    & 0      & 42,101  & 276.5       & 1           \\
                DS [MLST'24]~\cite{ds-fpga}$^a$    & 32        & $\le 75.9$\%        & 26 (130 ns)      & 903,284   & 434    & 358,754 & $\sim 200$  & 2           \\
                GNN [MLST'24]~\cite{ds-fpga}$^a$   & 32        & $\le 75.8$\%        & 32 (160 ns)      & 1,162,104 & 2,120  & 761,061 & $\sim 200$  & 3           \\
                \midrule
                \multicolumn{8}{c}{\textbf{PLF JSC (16 features)}}                                                                                                 \\
                \midrule
                Implementation                     & Particles & Accuracy $\uparrow$ & Latency [cycles] & LUT       & DSP    & FF      & \fmax [MHz] & II [cycles] \\
                \midrule

                HGQ [FPGA'26]~\cite{hgq}           & 64        & 82.4\%              & 26 (122.5 ns)    & 244,515   & 0      & 112,993 & 212.2       & 1           \\
                \textbf{HGQ-LUT (GNN)}             & 64        & 81.0\%              & 11 (28.7 ns)     & 39,765    & 0      & 19,637  & 383.6       & 1           \\
                HGQ [FPGA'26]~\cite{hgq}           & 64        & 80.7\%              & 9 (45.0 ns)      & 53,546    & 0      & 13,629  & 199.9       & 1           \\
                HGQ [FPGA'26]~\cite{hgq}           & 32        & 81.5\%              & 26 (119.8 ns)    & 238,255   & 0      & 116,039 & 217.1       & 1           \\
                \textbf{HGQ-LUT (GNN)}             & 32        & 80.3\%              & 10 (22.8 ns)     & 25,288    & 0      & 11,863  & 438.4       & 1           \\
                HGQ [FPGA'26]~\cite{hgq}           & 32        & 80.2\%              & 9 (45.5 ns)      & 48,343    & 0      & 10,012  & 197.7       & 1           \\
                GNN U4 [TECS'24]~\cite{llgnn}$^a$  & 50        & 80.9\%              & 130 (650 ns)     & 855k      & 8,945  & 201k    & $\sim 200$  & 100         \\
                GNN U5 [TECS'24]~\cite{llgnn}$^a$  & 50        & 81.2\%              & 181 (905 ns)     & 815k      & 8,986  & 189k    & $\sim 200$  & 150         \\
                GNN J4 [TECS'24]~\cite{llgnn}$^a$  & 30        & 78.4\%              & 58 (290 ns)      & 865k      & 8,776  & 138k    & $\sim 200$  & 30          \\
                GNN J5 [TECS'24]~\cite{llgnn}$^a$  & 30        & 79.9\%              & 181 (905 ns)     & 911k      & 9,833  & 158k    & $\sim 200$  & 150         \\
                GNN (FPL'22)~\cite{que2022opt}$^a$ & 50        & 80.4\%              & 2132 (10660 ns)  & 1515k     & 12,284 & 533k    & $\sim 200$  & 650         \\
                GNN (FPL'22)~\cite{que2022opt}$^a$ & 30        & 78.7\%              & 382 (1910 ns)    & 1158k     & 11,504 & 246k    & $\sim 200$  & 400         \\
                \midrule
            \end{tabular}

            \footnotesize
            \normalsize
        \end{threeparttable}
    \end{adjustbox}
    \begin{adjustbox}{width=0.85\textwidth,center=\textwidth}
        \begin{threeparttable}
            \begin{tabular}{l|ccccccc}
                \multicolumn{8}{c}{\textbf{Muon tracking}}                                                                                                  \\
                \midrule
                Implementation                           & Resolution $\downarrow$ & Latency [cycles] & LUT    & DSP   & FF     & \fmax [MHz] & II [cycles] \\
                \midrule
                \textbf{HGQ-Hybrid}                      & 1.90 mrad               & 5 (30.8 ns)      & 30,612 & 0     & 5,747  & 162.2       & 1           \\
                \textbf{HGQ-Hybrid}                      & 2.03 mrad               & 4 (24.5 ns)      & 19,627 & 0     & 3,157  & 163.1       & 1           \\
                \textbf{HGQ-Hybrid}                      & 2.28 mrad               & 4 (24.3 ns)      & 14,549 & 0     & 2,001  & 164.6       & 1           \\
                HGQ [FPGA'25]~\cite{hgq}                 & 1.90 mrad               & 8 (47.4 ns)      & 41,830 & 0     & 10,061 & 168.9       & 1           \\
                HGQ [FPGA'25]~\cite{hgq}                 & 2.03 mrad               & 6 (35.2 ns)      & 25,716 & 0     & 3,455  & 170.3       & 1           \\
                HGQ [FPGA'25]~\cite{hgq}                 & 2.38 mrad               & 5 (28.7 ns)      & 14,789 & 0     & 3,091  & 174.1       & 1           \\
                \texttt{QKeras} [NIMA'23]~\cite{tgc}$^a$ & 1.95 mrad               & 17 (106.3 ns)    & 37,867 & 1,762 & 8,443  & $>160$      & 1           \\
                \texttt{QKeras} [NIMA'23]~\cite{tgc}$^a$ & 2.04 mrad               & 13 (81.3 ns)     & 54,638 & 324   & 6,525  & $>160$      & 1           \\
                \texttt{QKeras} [NIMA'23]~\cite{tgc}$^a$ & 2.45 mrad               & 10 (62.5 ns)     & 28,526 & 24    & 2,954  & $>160$      & 1           \\
                \midrule
            \end{tabular}
            \vspace{-0.1cm}
            \footnotesize
            $^a$ \fmax is not provided, best estimated based on the target frequency/mentioning in the original works. \\
            \normalsize
        \end{threeparttable}
    \end{adjustbox}

    \vspace{-0.8cm}
\end{table*}

For experimental evaluation, we compare our HGQ-LUT method against prior works on five datasets, including the OpenML and CERNBox datasets for high-level feature (HLF) based jet substructure classification (JSC), and the particle level features (PLF) based JSC dataset from Zenodo~\cite{jet_dataset}, muon tracking for the Thin Gap Chamber (TGC) detector at the ATLAS experiment~\cite{tgc}, and cluster counting for particle identification (PID) at gas detectors based on drift chamber waveforms~\cite{cluster_counting_ML}.

Similar to the HGQ~\cite{hgq} work, we use a single training run to map out the Pareto frontier of validation accuracy versus estimated LUT utilization by sweeping the \texttt{beta} parameter with an exponential schedule during training. The \texttt{beta} parameter is a scalar factor that controls the trade-off between neural network performance and resource usage, where a larger \texttt{beta} leads to more compact models at the expense of performance. The initial and final \texttt{beta} values are the same as in~\cite{hgq}, which are \texttt{5e-7} to \texttt{1e-3} for HLF JSC, and \texttt{2e-8} to \texttt{3e-6} for both PLF JSC and TGC Muon Tracking. The optimizer used is Adam~\cite{adam}, and a cosine annealing with restarts learning rate schedule is used for all experiments. The best models on the Pareto frontier are then selected for test accuracy evaluation and hardware implementation. For all experiments, the validation set used for model selection during training is 10\% of the training set, which is always strictly disjoint from the test set. A single NVIDIA RTX 4090 GPU with an Intel Core i7-13700K CPU is used for all experiments.

Unless otherwise stated, we use {xcvu13p-flga2577-2-e} as the target device for all experiments, as it is the one expected to be used for future HL-LHC upgrades at CERN. For all experiments, we use the out-of-context, post-routing reports from Vivado 2025.1 for resource and \fmax\ measurements. All designs we evaluate are generated with the \texttt{da4ml} toolchain's Verilog flow. Since the pipelining heuristics used in \texttt{da4ml} are not yet optimized, we enabled global retiming for all designs during synthesis. No input clamping or other non-linear pre-processing was performed, nor needed, unless otherwise mentioned.

\subsection{Training Time Comparison}
We present the training time comparison between HGQ-LUT and prior works. We measure the training time per batch with the same hardware setup at the same batch sizes for each dataset. For all prior works, we use the official implementations provided by the authors, and measure the forward+backward+optimization step time in the training loop. Note that since the model sizes used for the same task may differ between different works, the time per batch may not be directly comparable. Nevertheless, we still consider it to provide a reasonable indication of the training efficiency of each method, and the results should be interpreted in conjunction with the model performance and resource usage achieved in the following sections. The results are shown in Tab.~\ref{tab:training_speed}, where we can see that HGQ-LUT achieves similar training time as plain HGQ~\cite{hgq}, which is two orders of magnitude faster than NLA~\cite{na}. Since the prior LAT methods were not evaluated on the other tasks in the literature, and no native multi-dimensional tensor support was provided in their implementations, we do not include the training time comparison for the other tasks here.

In the table, the training time reported for HGQ-LUT for the Muon Tracking task uses the hybrid architecture with both LUT-Dense and matmul-based layers, as later described in Section~\ref{sec:tgc}. If using only LUT-Dense layers, the training time per batch would be around 56.7 ms.

\subsection{JSC HLF}
\label{sec:hlf_jsc}

Both the OpenML and CERNBox variants of the dataset are used for high-level feature (HLF) based jet substructure classification (JSC). Each contains 16 features extracted from reconstructed jets in proton-proton collisions, with five-class labels corresponding to different originating particles of the jet: quark, gluon, W, Z, and top. The metric to evaluate the model performance is the overall classification accuracy, defined by the argmax of the model outputs.

\begin{figure}[htbp]
    \centering
    \includegraphics[width=0.49\textwidth]{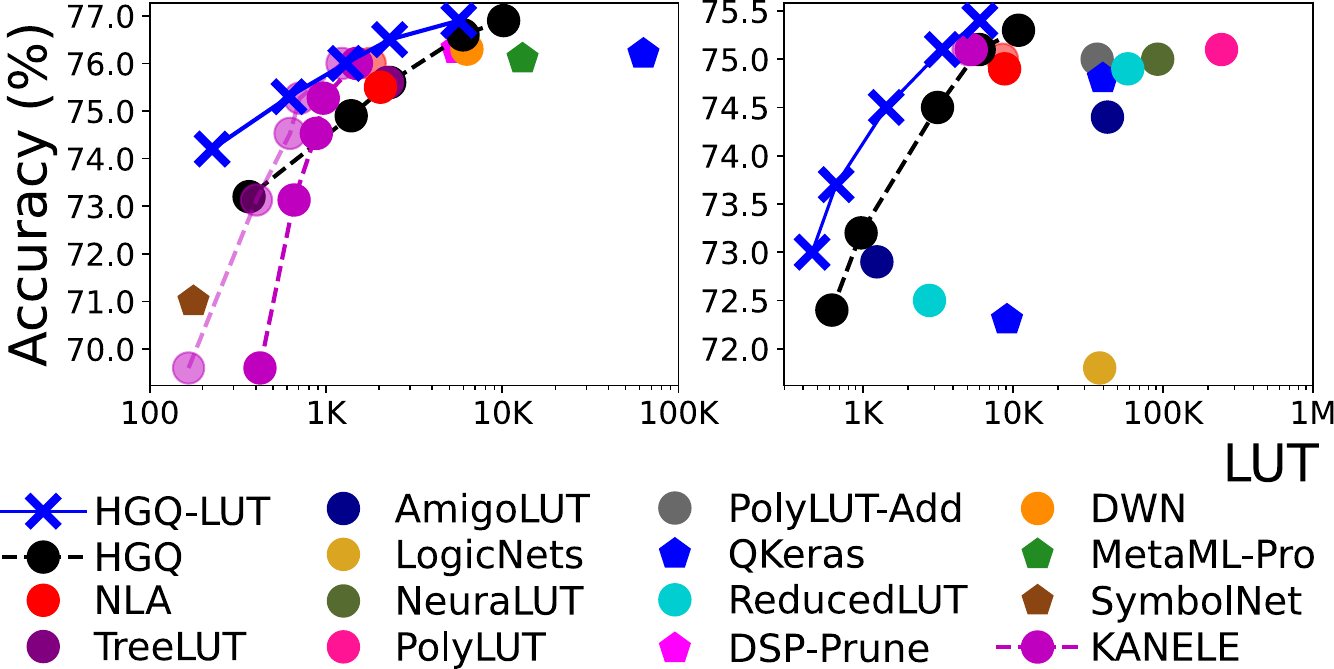}
    \vspace{-0.5cm}
    \caption{
        Accuracy versus LUT usage of the HLF JSC models on the OpenML and CERNBox datasets.
        Pareto frontiers are plotted by connecting the points from the same work when available.
        Designs with circle markers use only LUTs, and designs with pentagon markers use both LUTs and DSPs.
        Our designs are marked as "X" and use only LUTs.
        For NLA~\cite{na} and KANELE~\cite{kanele}, we show the reproduced/corrected results in solid points and the originally reported results in transparent points.
    }
    \vspace{-0.4cm}
    \label{fig:acc_lut_hlf}
\end{figure}

For this task, we use a two-layer architecture with 20- and 5- dimensional HGQ-LUT layers, with fused batch normalization enabled for the first layer. We compare the performance of the models trained with HGQ-LUT against various prior efforts, including vanilla HGQ, other neural networks based approaches, LUT-based approaches, decision forests, and symbolic models on the two HLF JSC datasets in Tab.~\ref{tab:hlf_jsc}, and visualize the accuracy versus LUT consumption in Fig.~\ref{fig:acc_lut_hlf} with both reported values from the original works and our reproduced results or numbers with corrections when necessary. In terms of accuracy and LUT-usage trade-off, HGQ-LUT outperforms all prior works on both datasets and achieves the best Pareto frontier, with a larger advantage in the low LUT region. HGQ-LUT also demonstrates significant latency reduction compared to the plain HGQ baseline, bringing latency down to a level similar to that of other LUT-based approaches, while still maintaining higher accuracy and lower LUT usage.

During our investigation, we found several issues in the prior works that lead to unfair comparisons that should be highlighted and corrected for the future research in this area. In particular, (1) for both NLA~\cite{na} and KANELE~\cite{kanele}, the authors {used the test set for validation directly during training and selected the best model over all epochs}. (2) The design {skipped the required input clamping in the hardware implementation}. This operation has an overhead of a few hundred LUTs with minor impact on latency. While this was also done in earlier works such as~\cite{hls4ml, qkeras, qkeras-xtre-q}, it was negligible due to the large firmware sizes. However, since the reported utilization can be lower than 2k LUTs in NLA and KANELE, the overhead becomes significant. We also notice that (3) the works used {unregistered outputs of the top module for out-of-context synthesis with global retiming} enabled, leading to missing timing paths in the reports and hence over-optimistic \fmax estimations. (4) {NLA used un-seeded shuffling on the OpenML dataset}, leading to $\sim80$\% of the samples used for test accuracy reporting were used previously in training. These findings can be validated independently with the official implementations provided by the authors~\cite{na-code,kanele-code}. For NLA, we used the official implementations provided with the above issues address (10\% validation split, proper seeding, input clamping, and registered outputs) and reproduced the results. For KANELE~\cite{kanele}, due to time constraints, we only estimate the LUT usage overhead from (3) to be the same as that of NLA~\cite{na}, and include the estimated LUT consumption with the overheads in LUTs added to the reported results, while keeping the potentially overestimated accuracies and \fmax values as-is.

\subsection{JSC PLF}

We further evaluate the performance of HGQ-LUT on the particle-level feature (PLF) based JSC dataset from Zenodo~\cite{jet_dataset}, which contains low-level, per-particle features of reconstructed jets in proton-proton collisions. Each sample contains up to 200 particles, each with 16 features, and the same five-class labels as in the HLF JSC datasets. When less than 200 particles are present, zero-padding is used. The inputs are hence arrays of shape ($N$, $F$), where $N$ is the number of particles and $F$ is the number of features per particle. In this work, we evaluate the performance of HGQ-LUT with $N=32,F=16$, $N=64,F=16$, and $N=32,F=3$ variants of the dataset. For the $F=3$ variants, only the three kinematic features, $p_T$, $\eta$, and $\phi$ are used for each particle, and we remove all particles with $p_T < 2$ GeV to simulate the effect of detector thresholds, following~\cite{ds-fpga}. The metric to evaluate the model performance is the same as the HLF JSC task, namely the overall classification accuracy defined by the argmax of the model outputs.

Due to the high input dimension and complexity of the dataset, prior LUT-based efforts are unable to handle this dataset directly. We use the same model architecture as in the JEDI-Linear~\cite{jedi-linear} work, but replacing all EinsumDense layers with our LUT-Dense layers. Since the original hidden dimension of 64 in the EinsumDense layers in the graph architecture is found to lead to excessive training time, we further reduce all hidden layer dimensions to 8 in our experiments. The results are compared against various prior efforts in Tab.~\ref{tab:plf_jsc_tgc}, showing that HGQ-LUT is able to achieve superior latency and resource efficiency compared to the vanilla matmul-based HGQ or other neural network based approaches, while lagging slightly behind in terms of absolute maximum accuracy achieved, albeit only by less than 1.5\%, potentially due to the reduced model dimensionality.

\subsection{TGC Muon Tracking}
\label{sec:tgc}

We compare the performance of HGQ-LUT on the TGC muon tracking dataset from Ref.~\cite{tgc}, which contains simulated hits from muons passing through the TGC detector at the ATLAS experiment. Each sample contains $7\times50$ bits extracted from the detector hits, with a regression target of the incident muon angle. The metric to evaluate the model performance is the mean squared error between the predicted and true angles with a cut-off at $30$ mrad~\cite{tgc}.

\begin{figure}
    \centering
    \includegraphics[width=0.35\textwidth]{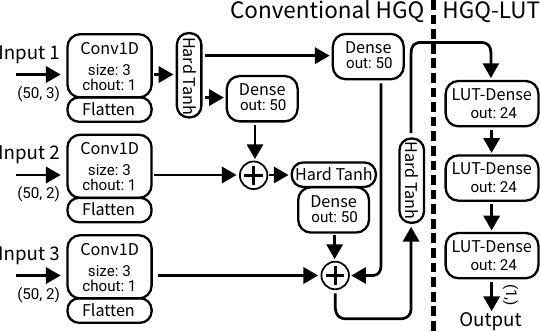}
    \caption{The hybrid architecture used for TGC Muon Tracking task with matmul-based dense layers (Plain HGQ) and LUT-Dense layers (HGQ-LUT).}
    \label{fig:tgc-arch}
    \vspace{-0.5cm}
\end{figure}

We found that using only LUT-Dense layers for this task would lead to significant accuracy degradation, potentially due to the high input dimension with low information content per feature: since each input is binary and we require L-LUTs to have only one input, the use of LUT-Dense-like layers may be inefficient for layers where low information content per feature is present. By using a hybrid architecture, we are able to achieve good accuracy while still benefiting from the low-latency inference of the LUT-Dense layers. Hence, for this task, we use a hybrid architecture with both LUT-Dense and matmul-based dense layers, where the feature-extraction part is realized with conventional dense layers, and the final dense output head is realized with LUT-Dense layers to achieve the best latency and resource efficiency, as shown in Fig.~\ref{fig:tgc-arch}.

With the hybrid architecture, we are able to achieve the same accuracies as the vanilla HGQ baseline with approximately 1/3 lower latency and 1/4 lower LUT utilization, as shown in Tab.~\ref{tab:plf_jsc_tgc}, demonstrating the advantage of HGQ-LUT for low-latency, resource-efficient inference on FPGAs.

\subsection{Particle Identification for Gaseous Detectors}

We evaluate our lookup-table (LUT)-based method using a drift chamber waveform dataset~\cite{cluster_counting_data} from Ref.~\cite{cluster_counting_ML} designed for particle identification (PID) via cluster counting at the proposed Circular Electron Positron Collider (CEPC)~\cite{cepc-tdr-18,cepc-tdr-24} experiment. Each input of the dataset is a simulated waveform sampled at 1.5 GSa/s with a total length of 3000 samples. Each sample in a waveform has a ground-truth label of whether it corresponds to the arrival of a primary ionization cluster, and the goal is to count the total number of primary ionization clusters in each waveform that is needed for PID.

The waveforms in the dataset represent the idealized continuous output of the front-end readout. In our evaluation, we digitize the waveforms to 12 bits fixed-point format \texttt{ap\_ufixed<3,9>}. As the frontend analog-to-digital converter in real hardware implementations would have a limited dynamic range, we include a pre-processing step to clamp the input waveforms to [0, $2^3-2^{-9}$] before digitization that is \emph{not realized in the hardware implementation}.

The dataset contains two classes of test samples, corresponding to two particle species, kaons and pions, each simulated under seven different kinematic conditions. The performance is evaluated per condition using the \emph{separation power} ($S$) as defined in Ref.~\cite{cluster_counting_ML}: $S = \frac{\mu(N_k) - \mu(N_p)}{\left[\sigma(N_k) + \sigma(N_p)\right]/2}$,
where $\mu(N)$ and $\sigma(N)$ are the mean and standard error of $N$ evaluated over samples in each condition, and the subscripts $k$ and $p$ denote kaons and pions, respectively.

\begin{figure}
    \centering
    \includegraphics[width=0.495\textwidth]{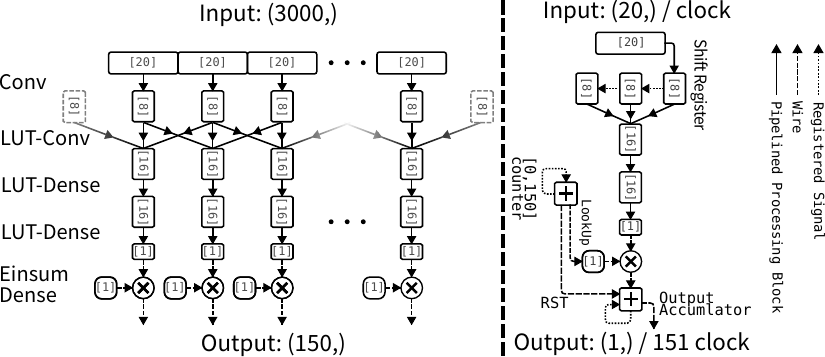}
    \caption{The overall network architecture used for the CEPC gas detector PID task during training (left) and FPGA-based inference (right).}
    \label{fig:cepc_architecture}
\end{figure}

Since the data acquisition system is limited to about 256 bits per cycle and each sample uses 12 bits, we assume 20 samples are fetched in parallel per cycle. This gives an initiation interval of 150 clock cycles plus 1 cycle of synchronization overhead per waveform at 75 MHz. The network is trained to regress the number of primary ionization clusters within each 20-sample window from 60 input samples, and the final count is obtained by accumulating the outputs over all windows. Since the inputs have a large bit-width, we find that feeding them into the LUT-layers directly leads to excessive area usage. Instead, we first use one conventional convolutional layer with matmul-based operations to project each 20-sample patch input into a lower dimension with 8 features, followed by three LUT-Conv layers to obtain a resource-efficient model. At the end of the network, we use an extra time-dependent lookup-table to weight the importance of the outputs at different time steps before the accumulation. The overall architecture for training and FPGA-based inference is shown in Fig.~\ref{fig:cepc_architecture}.

\begin{figure}
    \centering
    \includegraphics[width=0.45\textwidth]{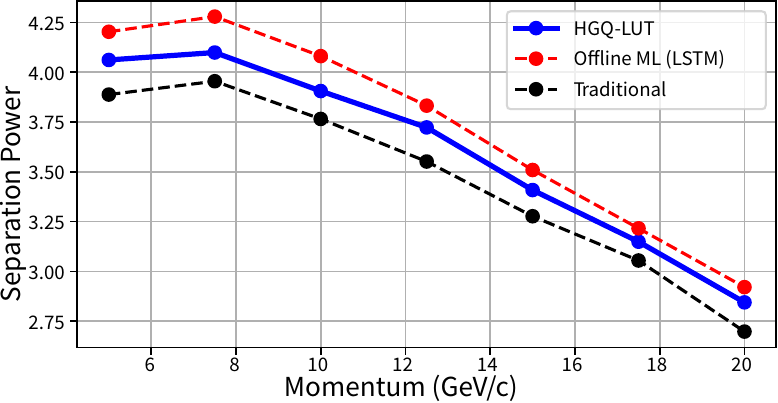}
    \vspace{-0.2cm}
    \caption{
        Separation power at different particle momenta for the HGQ-LUT model compared to the prior works.
    }
    \vspace{-0.55cm}
    \label{fig:cepc}
\end{figure}

We directly train a single model with a fixed \texttt{beta} value of \texttt{1e-7} since the objective is to achieve the best possible separation power within a LUT budget of under 10k LUTs for embedded FPGA deployment. The resulting separation power at different particle momenta is shown in Fig.~\ref{fig:cepc}, while using 6813 LUTs, 1 DSP, no BRAMs, 903 FFs, with a \fmax\ of 122.1 MHz, a latency of 154 and II of 151 cycles from out-of-context place and route reports. We show that with HGQ-LUT, we achieve separation power superior to the traditional offline reconstruction method in the online FPGA deployment setting, albeit still below the large LSTM model from Ref.~\cite{cluster_counting_ML} that is impractical for hardware deployment.

\section{Conclusion}

This paper proposes HGQ-LUT, a novel LUT-based neural network architecture and training method for efficient FPGA implementation. By introducing the new LUT-Dense and LUT-Conv layers into the \texttt{HGQ} and \texttt{da4ml} frameworks, we enable efficient training and streamlined hardware development of LUT-based neural networks. Extensive experimental evaluation on various high-energy-physics-related tasks demonstrates that HGQ-LUT achieves state-of-the-art accuracy and resource efficiency trade-offs. Simultaneously, HGQ-LUT significantly reduces training time compared to prior LUT-based training methods by over 100 times, enabling the practical use of LUT-based neural networks for previously intractable high-dimensional tasks.
HGQ-LUT not only opens up new possibilities for deploying efficient neural networks on FPGAs for real-time data processing in resource- and latency-constrained applications, it also provides a novel basis for next-generation LAT tools for machine learning acceleration.
Future work will explore extending HGQ-LUT to broader model architectures, such as transformers~\cite{laatu2025sub, zheng2026jetformer}, and developing more efficient mappings across different FPGA resources~\cite{ma2025cd} and platforms.

\vspace{0.3cm}
\noindent \textbf{Acknowledgement.}
Partial support from the United States DoE (grant numbers DE-SC0011925, DE-FOA-0002705), NSF (grant numbers PHY240298, PHY2117997), United Kingdom EPSRC (grant numbers UKRI256, EP/V028251/1, EP/N031768/1, EP/S030069/1, and EP/X036006/1), KIAT, Intel, and AMD is acknowledged.

\footnotesize
\bibliographystyle{IEEEtran}

\bibliography{biblio}

\balance

\end{document}